# Spin wave mode conversion in an in-plane magnetized microscale T-shaped YIG magnonic splitter

Takuya Taniguchi[1,2], Jan Sahliger[2], and Christian H. Back[2]

[1]Institute of Multidisciplinary Research for Advanced Materials, Tohoku University, Sendai 980-8577, Japan

[2]Fakultät für Physik, Technische Universität München, Garching 85748, Germany

[Abstract]

As one of the fundamental magnonic devices, a magnonic splitter device has been proposed and spin wave propagation in this device has been studied numerically and experimentally. In the present work, we fabricated a T-shaped magnonic splitter with 6 μm-wide three arms using a 100 nm-thick yttrium iron garnet film and, using time-resolved magneto-optic Kerr microscopy, observed that spin waves split into both, the vertical and the horizontal direction at the junction. Analyzing the results, we found that spin wave modes are converted into another during the splitting process and the splitting efficiency is dominantly dependent on the 1$^{st}$ order of incoming spin waves.

[Main article]

Magnetic moments form spin waves (SWs) as fundamental collective excitations in a magnetically ordered system. When SWs propagate in a magnetic media, they convey spin information in space, which can be exploited in so-called magnonic devices [1-3]. Since the propagation is not accompanied by Joule losses, magnonic devices have been drawing attention as potential low-energy-consuming data-processing devices. The proposed magnonic devices are generally based on the wave properties of SWs [4-9], which potentially enable multi-valued computation leading to high-density integrated circuits in contrast to CMOS-based devices. To design SW-based circuits, it is necessary to understand how magnonic splitter devices work. As one of the simplest magnonic splitter structures, SW propagation in T-shaped devices has been investigated and it was experimentally studied how SWs propagating through one branch of a T-structure are converted to different SW modes propagating in the two other branches [10-12]. However, although these works showed indeed that SWs can be split using T-shaped devices, the details of the SW conversion and of the transmission processes in T-shaped magnonic splitter remain concealed. To further reveal the properties of T-shaped magnonic splitters, we previously investigated SW conversion processes using micromagnetic simulations and found that even higher order SW modes can be selectively excited in properly designed structures [13]. In the present work, we experimentally study the SW conversion processes and discuss their dependencies on the applied magnetic field.

To experimentally investigate SW propagation, we designed a T-shaped spin wave splitter utilizing photolithography (Fig.1(a)). The T-shaped device has 6 μm-wide and 100 μm-long three arms and it was fabricated by Ar ion etching a 100 nm thick yttrium iron garnet (YIG) film grown on gadolinium gallium garnet (GGG) by liquid phase epitaxy. In figure 1(c) and (d), we display the profile of the effective field and of the demagnetization factor's ratio $N_X/N_Z$ near the junction calculated using Mumax3 [14, 15]. Additionally, a 5 μm wide microstrip consisting of Cr (5 nm)/Au (200 nm) was attached to the device utilizing photolithography and e-beam evaporation. In a static external magnetic field applied along the y-direction, SWs were excited by the Oersted field generated by an rf electric current (2.4 GHz or 4.8 GHz) flowing through the microstrip, and SW propagation was observed by time-resolved polar magneto-optic Kerr effect microscopy (TR-MOKE) [16]. Since the wavevector of the propagating SWs and the direction of the static field are orthogonal to each other and both are aligned in the plane, the excited SWs are Damon-Eshbach SWs (DESWs) [17]. We present the typical results obtained at the condition of 2.4 GHz and 34.6 mT in Fig 2(a) and (b) and at the condition of 4.8 GHz and 106 mT in Fig. 2(c) and (d). Figure 2 displays the reflection of the laser and Kerr signals corresponding to the topography of the device and the z-component of the dynamic magnetization ($m_z$), respectively. One can see SWs propagating through the left arm ($x \lesssim -7$ μm) and at the junction ($-7$ μm $\lesssim x \lesssim 7$ μm) they are partially converted to SWs propagating along the y-direction in the transverse arm (6 μm $\lesssim y$). Furthermore, the incident DESWs are partially transmitted to SWs propagating along the x-direction in the right arm (7 μm $\lesssim x$). We would like to note that the SWs propagating in the transverse arm have magneto-static backward volume SW (MSBVW) character and the SWs propagating in the right arm have DESW geometry. Similar to our simulation results [13], we experimentally observed the excitation of multi-modes of MSBVWs. Moreover, in addition to the MSBVWs, multi-modes of transmitted DESWs were also observed even though we did not observe mode conversion from incident DESWs to transmitted DESWs in our previous numerical work.

For characterizing the observed SWs' properties, we extract the Kerr signal from each arm independently and fit the SW signals. Due to confinement, propagating SWs have energetically distinct eigenmodes depending on the standing waves formed along the width and thickness direction [10-13, 18, 19]. Since our film is thin, we fit the results using the equation below, which takes the standing waves formed only along the width direction into account:

$$m_z(p_{arm}, q_{arm}) = \sum_n A_{n,q=0}^{arm} \sin\left(\frac{n\pi}{w_{\text{eff}}^{arm}} \cdot p_{arm}\right) \sin(k_{n,q}^{arm} q_{arm} + \phi) \exp\left(-\frac{q_{arm}}{\Lambda_n^{arm}}\right) + Const.. \quad (1)$$

Here, $p_{arm}$ and $q_{arm}$ are the relative coordinates in each arm: $p_{arm}$ is along the width direction and $q_{arm}$ is along the longitudinal direction of each arm. Integer $n$ is the mode number defined by the number of antinodes of the standing waves, $A_{n,q=0}^{arm}$ is the SW amplitude at $q_{arm} = 0$, $w_{\text{eff}}^{arm}$ is the effective width of each arm considering the pinning condition of the standing waves at the edges [20, 21], $k_{n,q}^{arm}$ is the wavenumber of the propagating SWs, $\phi$ is the SW's phase at $q_{arm} = 0$ and at the time when the stroboscopic TR-MOKE images were taken, and $\Lambda_n^{arm}$ is the attenuation length of the SWs. Note that *arm* (=left, right, transverse) of each parameter indicates the parameter obtained from the SWs propagating in the corresponding arm of the device. In addition, the equation is completed by a constant offset in order to consider optical artifacts. As shown in Figure 2, the results are well fitted by taking into account SWs up to the 3$^{rd}$ order.

To verify the fitting results, we varied the static external magnetic field and repeated the fitting procedure to obtain the SW dispersion relation. Figure 3 displays the SW wavelength as a function of the effective magnetic field: $\lambda_{DESW}$ and $\lambda_{MSBVW}$ are the wavelengths of DESWs and MSBVWs, respectively. Note that the effective magnetic field is extracted from micromagnetic simulations [14]. The experimental results are supplemented by the dispersion relation obtained theoretically [22,23]:

$$f = \frac{\gamma}{2\pi}\sqrt{[\mu_0 H_{eff} + \mu_0 M_s(1 - p + l_{ex}^2 k^2)]\left[\mu_0 H_{eff} + \mu_0 M_s\left(p\frac{k_x^2}{k^2} + l_{ex}^2 k^2\right)\right]}, \qquad (2)$$

where $f$ is the SW excitation frequency, $\gamma = 28$ GHz/T is the gyromagnetic ratio, $\mu_0$ is the permeability of vacuum, $H_{eff}$ is the static effective magnetic field, $M_s$ is the saturation magnetization, $p = 1 - (1 - e^{-kt})/kt$, $t$ is the thickness of the magnetic film, $k^2 = k_x^2 + k_y^2$, and $l_{ex}$ is the exchange length [24]. Note that $k_{n,q}^{arm} = k_x(k_y) = 2\pi/\lambda_{DESW}(2\pi/\lambda_{MSBVW})$ and $n\pi/w_{eff}^{arm} = k_y(k_x)$ for propagating DESWs (MSBVWs). The experimental results satisfactorily agree with the theory, which indicates not only that the obtained fitting results are valid but also that the excited DESWs and MSBVWs are conventional SWs propagating in a magnetic micro-stub.

Next, to discuss the functionality of the T-shaped magnon splitter, we firstly estimate the amplitude of each SW propagating in each arm would be at the center of the T-junction ($q = q_c \Leftrightarrow (x,y) = (0,0)$, see figure 2) using the following equation:

$$A_{n,center}^{arm} = A_{n,q=0}^{arm} \exp\left(-\frac{q_c}{\Lambda_n^{arm}}\right). \qquad (3)$$

Subsequently, since the SW excitation efficiency using a microstripe depends on the wavelength of the SWs [25], we calculate the normalized amplitude using the 1$^{st}$ order of the incoming SW mode, which is the dominant SW mode excited at the antenna: $a_n^{arm} = A_{n,center}^{arm}/A_{n=1,center}^{left}$. Figure 4(a)-(d) show $a_n^{arm}$ as a function of the external magnetic field and one can find that $a_{n=1}^{right}$ and $a_n^{transverse}$ depend on the magnetic field. This behavior of $a_n^{transverse}$ can be explained by concerning the magnon gap and the mode selectivity of the SW conversion process in a T-shaped device [13]. According to the micromagnetic simulation study, when the wavelength of the incident DESW, $\lambda_{DESW}$, matches to $2w_{eff}^{transverse}/n_{transverse}$, MSBVWs having the mode number, $n_{transverse}$, start being excited. To evaluate the field dependence of $a_n^{transverse}$, we display $a_n^{transverse}$ as a function of $2w_{eff}^{transverse}/\lambda_{DESW}$ in Fig. 4(e), (f). For both conditions, $f = 2.4$ GHz and 4.8 GHz, $a_n^{transverse}$ increases when

$2w_{\text{eff}}^{transverse}/\lambda_{\text{DESW}}$ gets closer to $n_{\text{transverse}}$: e.g., $a_{n=1}^{transverse}$ and $a_{n=2}^{transverse}$ overcome 1 at $2w_{\text{eff}}^{transverse}/\lambda_{\text{DESW}} = 0.45$ and $0.81$ for $f = 2.4$ GHz, and at $0.80$ and $1.49$ for $f = 4.8$ GHz, respectively. We must address that $a_{n=2}^{transverse}$ is still lower than $a_{n=1}^{transverse}$ even when $2w_{\text{eff}}^{transverse}/\lambda_{\text{DESW}}$ is close to 2 for $f = 4.8$ GHz. Such behavior was also observed in the previous numerical study by increasing the width of the transverse arm [13]. Since the incoming SWs diffract at the junction and the diffracted plane wave contains the 1st order of MSBVW, $a_{n=1}^{transverse}$ can be high when the diffraction at the junction is not negligible. Moreover, we would like to note that $a_n^{transverse}$ can exceed 1 due to the ellipticity of the magnetization precession [16]. Since the external magnetic field is aligned along the y-direction, the dynamic magnetization projected onto the zx-plane shows an elliptical trajectory, which produces a dynamic demagnetization field in the zx-plane. As seen in figure 1(c), the ratio of the local demagnetization factors in zx-plane $N_X/N$ in the transverse arm is almost 3 times larger than in the horizontal arm. It indicates that even if only the half amount of magnons is converted from DESWs to MSBVWs, MSBVWs have larger $m_z$ than DESWs. Since the MOKE signal in the present study is sensitive to the z-component of the dynamic magnetization, $a_n^{transverse}$ can exceed 100%.

We subsequently discuss the field dependence of $a_n^{right}$ (Fig. 4(c), (d)). While $a_{n=1}^{right}$ varies in a non-negligible range (0.2 – 0.8), $a_{n=2,3}^{right}$ stays at an almost constant level: $a_{n=2}^{right} \sim 0.35$ and $a_{n=3}^{right} \sim 0.2$ at 2.4 GHz and $a_{n=2}^{right} \sim 0.2$ at 4.8 GHz. It indicates that the T-junction converts the propagating mode of DESWs from $n = 1$ to the higher order mode at almost any measured condition. The higher-order excitation can be understood by considering the spatial distribution of the SW amplitude at the junction [26,27]. The spatial distribution of the SW amplitude may be due to (i) anisotropic SW scattering and (ii) anisotropic SW attenuation: (i) At the junction, the effective magnetic field locally varies due to the shape anisotropy (Fig. 1(b)). Since the junction has an asymmetric shape, the variation of the effective field leads to anisotropic scattering of the SWs, which results in the asymmetric profile of the DESW amplitude. (ii) When the incident DESWs arrive at the junction area, DESWs are converted to MSBVWs which attenuate along the y-direction. The attenuation of MSBVWs also induces a y-dependent DESW amplitude. As a consequence, the 2nd and 3rd order DESWs are excited after the DESWs pass through the junction. The mode conversion during the transmission process was not observed in the reported numerical work [13] possibly because (i) the scattering source, i.e. the variation of the effective field, was relatively large compared to the junction area so that an asymmetric SW amplitude was not generated and/or (ii) the attenuation length of transversally propagating SWs was much longer than the width of the horizontal branches so that the SW amplitude did not sufficiently vary along the width direction.

We finally attempt to have a deeper insight into the mode dependence on the conversion process and the transmission process. As discussed above, the anisotropic distribution of the SW amplitude at the junction does not efficiently generate the 1st order of DESWs. Hence, $a_{n=1}^{right}$ reflects the amplitude of the 1st order of the incoming DESWs and the loss due to the scattering and the generation of MSBVWs. Since $a_{n=2,3}^{right}$ do not vary over the error range and $a_{n=2,3}^{left}$ appear only at some conditions, the magnetic field dependence of $a_{n=1}^{right}$ basically reflects the generation of MSBVWs and the forward/backward scattering between the 1st order of the in-coming and out-coming DESWs. Comparing $a_{n=1}^{right}$ and $a_{n=1}^{transverse}$ in the same magnetic field range (32.3 – 37.7 mT at 2.4 GHz and 105.6 – 108.5 mT at 4.8 GHz), we find that $a_{n=1}^{right}$ tends to increase and $\sum a_n^{transverse}$ tends to decrease with the external magnetic field. The inverse field dependence indicates that the SW conversion from DESWs to MSBVWs dominantly originates from the 1st order of the DESWs. Note that the field dependent $a_{n=1}^{right}$ is not simply opposite to the field dependent $\sum a_n^{transverse}$. The possible mechanism behind this is as follows. Incoming DESWs are scattered at the T-junction due to the local effective field variation. Referring to the study of SW scattering [28], the ratio of the amplitude of the backscattered SWs to the one of the forward scattered SWs depends on the magnetic field. It indicates that $a_{n=1}^{right}$ contains the magnetic field dependent scattering because $a_{n=1}^{right}$ is estimated using $A_{n=1,\text{center}}^{left}$ that is the result obtained after the scattering event. Thus, $a_{n=1}^{right}$ contains two functions of the magnetic field, corresponding to the scattering and the conversion, and it results in the incomplete opposite behavior of $a_{n=1}^{right}$ and $\sum a_n^{transverse}$.

In conclusion, we fabricated a T-shaped magnonic splitter from a YIG thin film and observed SW propagation in the device utilizing TR-MOKE. DESWs were excited by an rf-current at an antenna and DESWs were converted to MSBVWs and transmitted to other DESWs at the T-junction. Estimating the conversion efficiency, the mode selectivity of MSBVWs is experimentally observed as numerically investigated. Moreover, we revealed that the MSBVWs are excited dominantly by the 1st order of DESWs. Additionally, it is found that the mode conversion occurs also during the transmission process. The mode conversion during the transmission is understood by the spatial distribution of SWs at the T-junction caused by anisotropic SW scattering and attenuation. This result implies that, if one wants to convey the original spin information through a T-shaped magnonic splitter, the splitter must be designed small enough in order to avoid anisotropic scattering events causing SW mode conversion during the transmission process. On the other hand, magnonic splitters are suggested to be designed on large scales if the conversion of the transported spin information is of interest.

[Acknowledgement]

We acknowledge financial support by JSPS Overseas research fellowships.

[References]

[Figures]

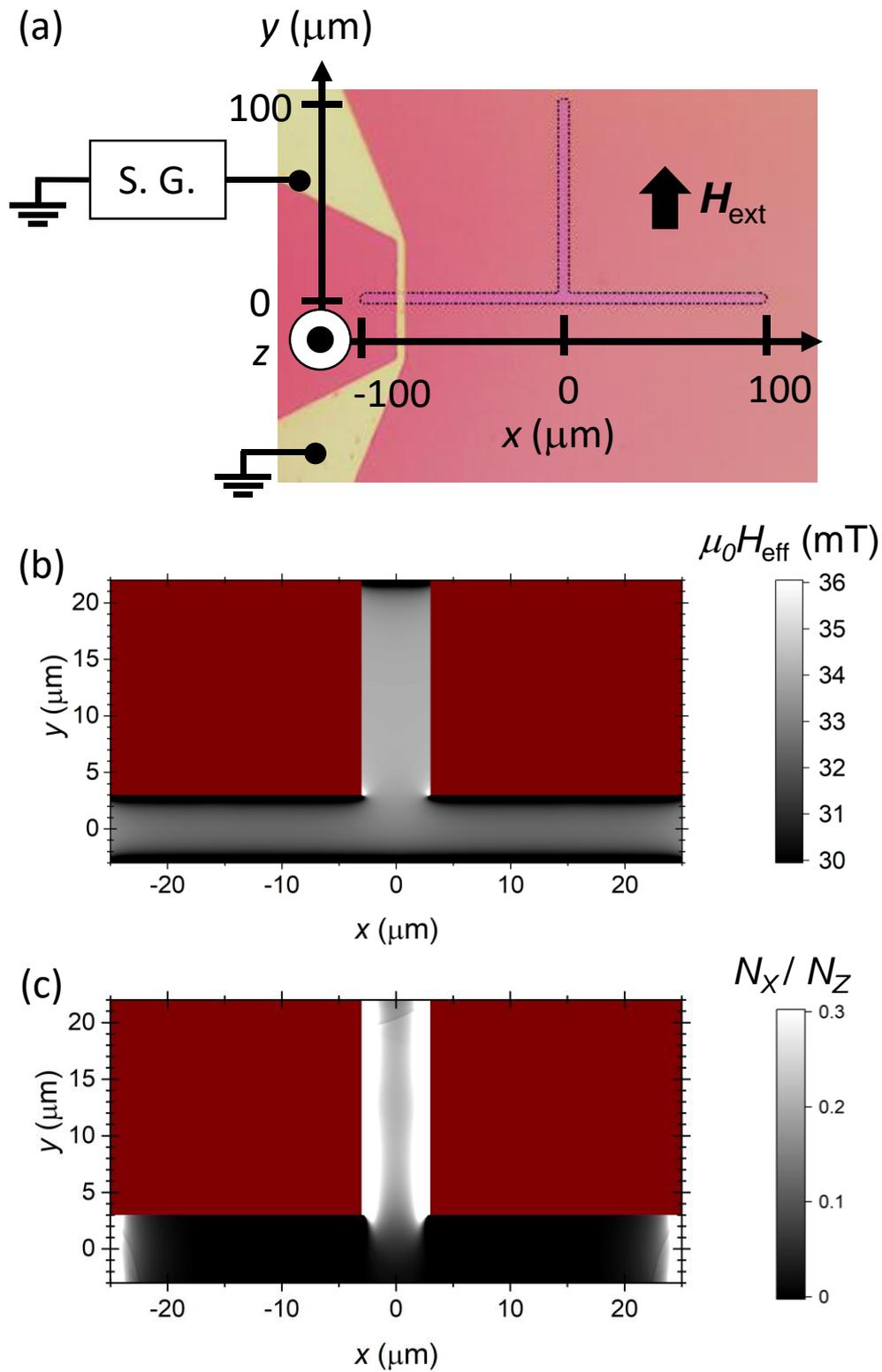

Fig.1 (a) Schematics of the experimental setup with an optical microscope image. The area surrounded by the broken lines indicates the magnonic splitter made of YIG. An rf current is applied through the gold microstrip (yellow area) using a signal generator (S.G.). (b) Internal effective magnetic field and (c) the demagnetization factor ratio $N_X/N_Z$ obtained by micromagnetic simulation [16]. The applied external magnetic field is (b) 34 mT and (c) 70 mT and the dark red regime indicates the space outside the device.

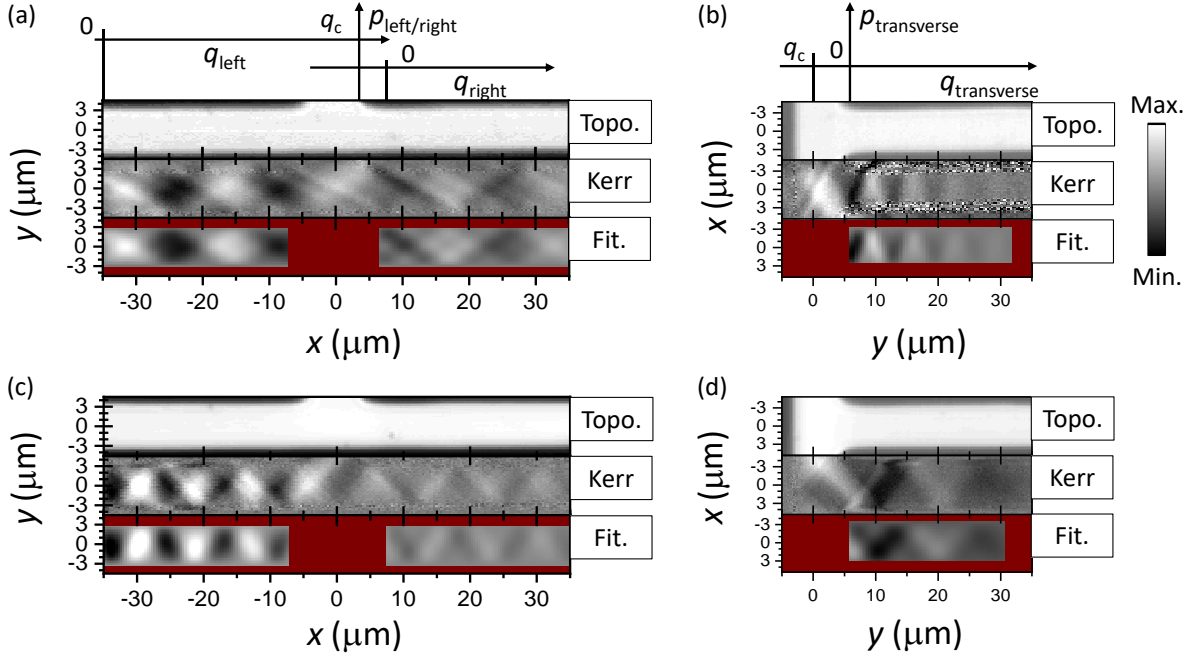

Fig. 2 Topography obtained by reflection (Topo.) and MOKE signals (Kerr) as well as fitting results (Fit.) from the experiment using SW the excitation condition $(f, \mu_0 H_{ext}) = (2.4\text{ GHz}, 34.6\text{ mT})$ for (a) and (b) and $(4.8\text{ GHz}, 106\text{ mT})$ for (c) and (d). We display each contour plot taken (a, c) from the horizontal branches and (b, d) from the transverse branch. The corresponding relative coordinate, $p_{arm}$, $q_{arm}$ (arm: the left, right, or transverse arm used for the fitting), and $q_c$ for each coordinate are also described. The dark red regime is the area, which we did not use for the analysis. Note that the contrast of each panel is set so that the waveforms are clearly visible..

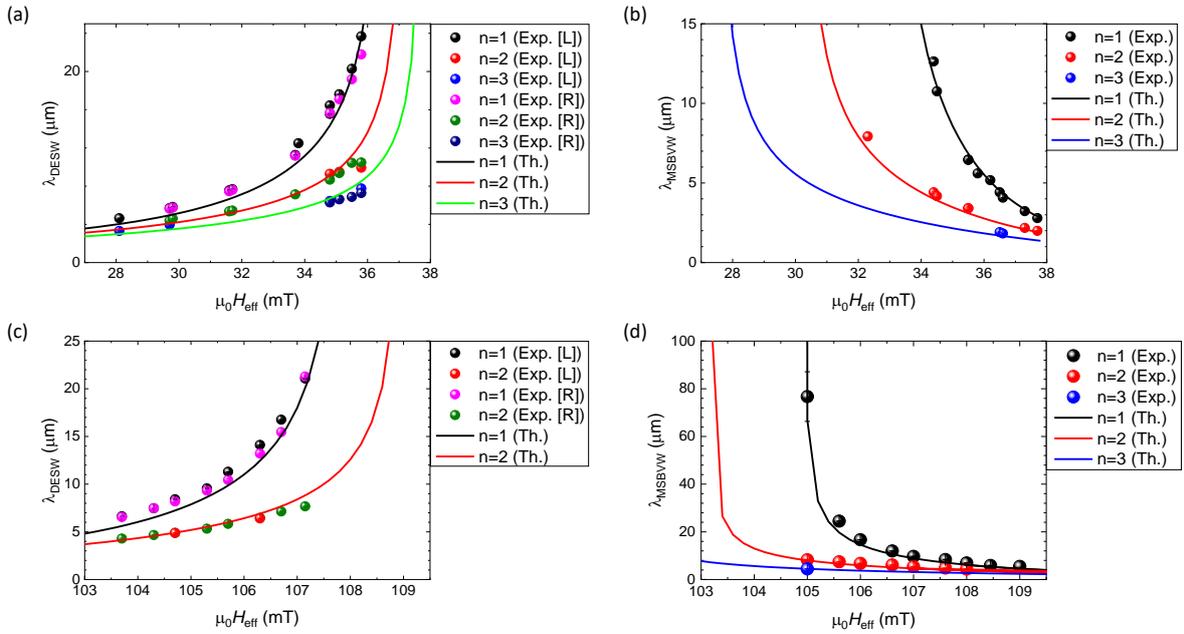

Fig.3 Dispersion relation obtained from the experiments and calculated from theory. (a,b) 2.4 GHz and (c,d) 4.8 GHz are used as the excitation frequencies for (a,c) DESWs and (b,d) MSBVWs. [L] and [R] in the legends indicate the results obtained from the left side and the right side of the horizontal branches, respectively.

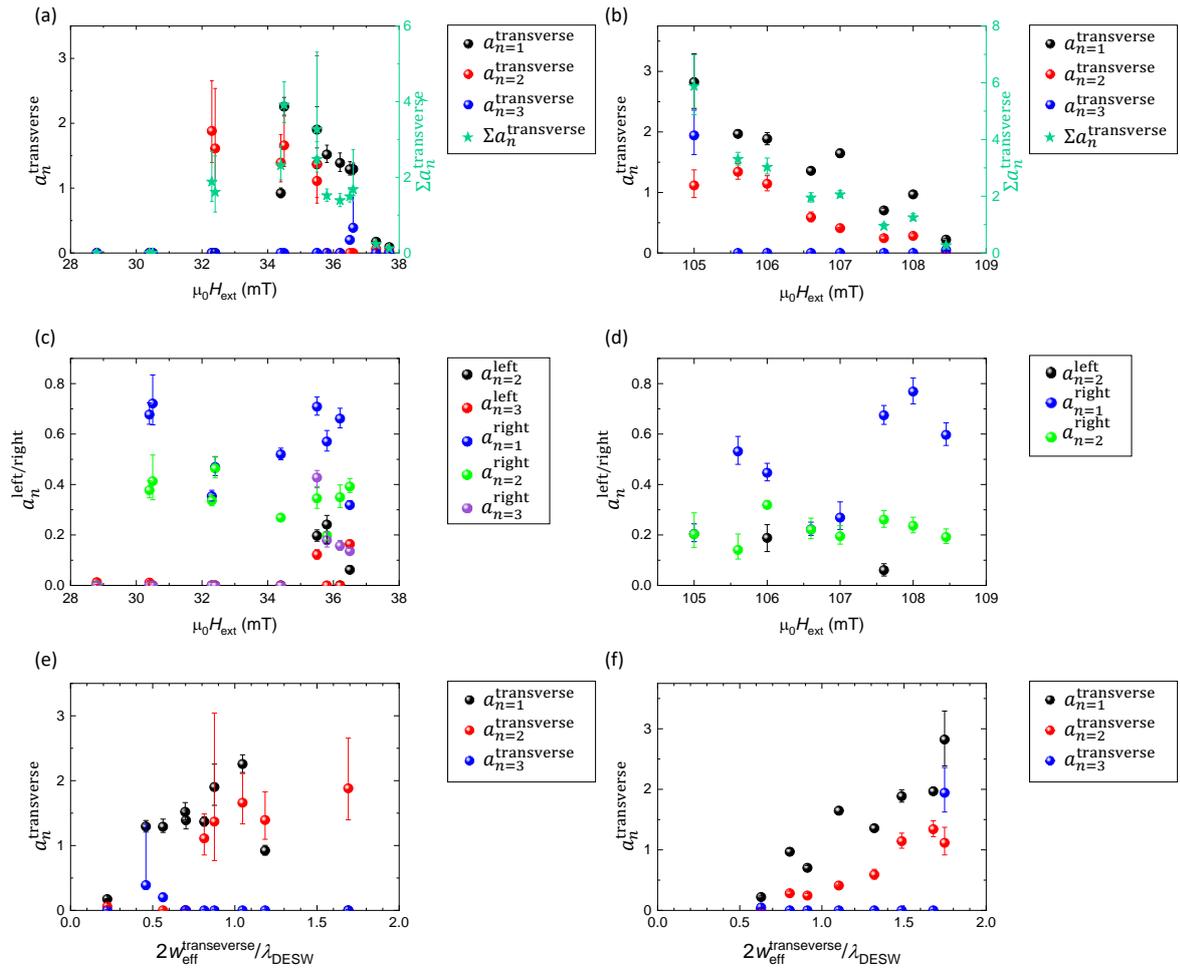

Fig.4 (a,b) $a_n^{\text{transverse}}$ and $\sum a_n^{\text{transverse}}$ and (c,d) $a_n^{\text{left/right}}$ as a function of the applied magnetic field. (e,f) $a_n^{\text{transverse}}$ as a function of $2w_{\text{eff}}^{\text{transverse}}/\lambda_{\text{DESW}}$. The SW excitation frequencies were (a,c,e) 2.4 GHz and (b,d,f) 4.8 GHz.